\documentstyle[12pt]{article}
\def\be{\begin{equation}}
\def\ee{\end{equation}}
\begin{document}

\begin{center}

{\bf Proposed Measurement of an Effective Flux Quantum in
the Fractional Quantum Hall Effect}

\vspace{1cm}

J.K. Jain$^{1}$,  S.A. Kivelson$^{2}$, and D.J. Thouless$^3$

\vspace{1cm}

1. {\em Department of Physics, State University of New York at Stony
Brook, Stony Brook, New York, 11794-3800}

2. {\em Department of Physics, University of California at Los
Angeles, Los Angeles, California 90024}

3. {\em Department of Physics, University of Washington,
Seattle, Washington, 98195}

\end{center}

\vspace{2cm}

\noindent

We consider a channel of an incompressible fractional-quantum-Hall-effect
(FQHE) liquid containing an island of another FQHE liquid. It is predicted
that the resistance of this channel will be periodic in the flux
through the island, with the period equal to an odd integer multiple
of the fundamental flux quantum, $\phi_{0}=hc/e$. The multiplicity
depends on the quasiparticle charges of the two FQHE liquids.

\pagebreak

Since the seminal works of Laughlin \cite {laughlin}
and Halperin \cite {halperin},
it has been recognized that the elementary excitations (quasiparticles)
in the fractional quantum Hall effect (FQHE) \cite {fqhe}
have fractional charge and obey fractional statistics. These
fractional quantum numbers essentially follow from the
incompressibility at fractional filling factors, and their values can
be determined from rather general principles \cite {su}. For the
principal FQHE liquids at filling factors \cite {ff}
\be
\nu_{n}\equiv \frac{n}{2n+1}\;\;,
\ee
the charge of a quasihole is
\be
e_{n}=\frac{e}{2n + 1} \;\;,
\label{charge}
\ee
while its  statistics is
\be
\theta_{n}= \frac{2n-1}{2n+1}\;\;,
\label{statistics}
\ee
defined so that an exchange of two quasiholes produces a phase
factor of $e^{i\pi\theta}$ \cite {general}. It has been argued
that the fractionally quantized Hall resistance itself is a
measurement of the charge of the quasiparticles \cite {laughlinbook},
but, on the other hand, the Hall resistance is a property of the
condensate and therefore does not {\em directly} probe the
excitations \cite {kp}. The observation of the `hierarchy
fractions' has been cited as evidence for the fractional
statistics of quasiparticles \cite {laughlin3},
but it is clear that all fractions {\em can be}
understood without reference to quasiparticles at all \cite
{jain89}. Several experiments have reported evidence for the fractional
charge \cite {simmons}. However, their theoretical interpretation is either not
unique, or not completely understood.
A {\em definitive} and {\em direct} observation of the fractional
charge or the fractional statistics of the quasiparticles is therefore
lacking.

In order to illustrate the basic conceptual difficulty with the
measurement of the fractional charge, consider the Aharonov-Bohm
(AB) geometry in Fig.1a. In the FQHE regime, the current is
carried by fractionally charged quasiparticles, so it is tempting to
expect that the properties of the system, such as the resistance,
will be periodic in the flux with period $\phi_{0}^*=hc/e_{n}$,
in analogy with the argument of Byers and Yang (BY) \cite {by}.
However, in any true AB geometry, the period must always be
$\phi_{0}=hc/e$.
The reason is that while the quasiparticles may provide an {\em
effective} description, the fundamental particles are still
electrons \cite {kr}.  In fact, periods {\em greater} than $\phi_{0}$
are ruled out by the BY argument (while smaller periods are, of course,
possible and do occur, e.g. in the case of superconductors).

In this Letter, we consider a resonant
tunneling experiment and predict that, under certain conditions, the
resistance will exhibit approximate periodicity in flux with
period equal to an odd integer multiple of $\phi_{0}$.
An observation of this periodicity should provide
direct and unambiguous evidence of the existence of
fractional quantum numbers in the FQHE. There have been other
proposals for the observation of the fractional quantum numbers \cite
{kivelson}, but they deal with {\em non-equilibrium} situations. The
experiment proposed in the present work, on the other hand, probes
an {\em equilibrium} property of the system.

We consider the geometry of Fig.1b, in which a (narrow) channel of
$\nu'=p'/q'$ FQHE liquid (where $p'$ and $q'$ are relatively prime
integers) contains an
island of area $A$ of the $\nu=p/q$ FQHE liquid.
This could be produced experimentally by creating a gentle potential
hill or valley with the help of an external gate.
The chemical potential at the edges of the sample
is assumed to be fixed externally.  The BY argument clearly does
not apply in this situation, since electrons occupy the
entire sample. It is possible for a
quasiparticle to tunnel from one edge of the channel to the other,
which is actually a tunneling between two many-body
configurations, one in which the quasiparticle is on one edge, and the
other in which it is on the other. The tunneling
amplitude determines the longitudinal resistance, as was shown
in a Landauer-type formulation of the QHE \cite {streda}. The longitudinal
resistance exhibits peaks whenever there is {\em resonant}
tunneling from one edge of the sample to the other through a
quasi-bound state on the potential island \cite {jain88}.
The main conclusion of this work is that successive
peaks occur when the flux through the island changes by
\be
j\phi_{0}=\frac{q}{s}\phi_{0}\;\;,
\label{tpd}
\ee
where $s$ is an integer, equal to
the highest common factor of $q$ and $q'$. Since $q$ and $q'$ are, in
general, odd integers, $j$ is also an odd integer. Note that $j$
depends only on $q$ and $q'$, i.e.,
only on the quasiparticle charges of the two FQHE liquids.

To give the simplest derivation of this result, let us change
the flux through the $\nu=p/q$ island liquid in a way that
no quasiparticles (quasiholes or quasielectrons) are created in the bulk.
This can be achieved by spreading the additional flux over a
sufficiently large area of the island.
The additional flux $j\phi_{0}$ contracts the island liquid, so that an
additional charge $jep/q$ is required to restore the edge
of the island FQHE liquid to its original state.
Since the charge must be supplied
by $j'$ quasiparticles of the channel ($\nu'$) FQHE liquid, we must have
\be
j\frac{p}{q}=j'\frac{1}{q'}\;\;,
\ee
which leads to the period $j\phi_{0}$ given by Eq.~(\ref{tpd}).
In particular, if $\nu=0$, i.e. if the island is charge free, the
period is $\phi_{0}$, since the channel FQHE liquid can return to its
original state by the transfer of $p'$  quasiparticles from the outer edge
to the inner edge. (Thus, $\nu=0$ is to be
interpreted as $\nu=0/1$ for the purpose of Eq.~\ref{tpd}.) This is
equivalent to a gauge transformation of the original wave function.

Let us now give a more microscopic description, which
takes account of the internal structure of the various FQHE liquids.
We use the framework of the composite fermion (CF)
theory \cite {jain89}, in which the
the wave function of the $\nu_{n}$ FQHE liquid is given by
\be
\chi_{n/(2n+1)}=\prod_{j<k}(z_{j}-z_{k})^{2}\Psi_{n}\;\;,
\ee
where $\Psi_{n}$ is the wave function of $n$ filled Landau levels
(LLs), and $z_{j}=x_{j}+iy_{j}$ denotes the position of the $j$th
electron. Consider the situation when the island FQHE liquid is
$\nu_{n-1}$ and the channel FQHE liquid is $\nu_{n}$.
This state corresponds to an integer quantum
Hall effect (IQHE) state which has $n$ filled LLs everywhere except
in an island where the filling factor is
$n-1$. An integer number ($K$) of electrons have been removed from the
$n$th LL to create the island \cite {semiclassical}.
In the IQHE state $\Psi$, each hole has an excess charge $e$ associated
with it.  Upon multiplication by the Jastrow factor,
$\prod_{j<k}(z_{j}-z_{k})^{2}$, which converts each electron into a
CF, each hole in the $n$th LL of $\Psi$ becomes a quasihole of the
$\nu_{n}$ liquid, with an excess charge $e_{n}=e/(2n+1)$ associated
with it \cite {charge}. Therefore, for $K$
quasiholes, there is a net deficiency of charge $Ke_{n}$ in the island
region. This deficiency is related to the difference between the
filling factors outside and inside the island as:
\be
Ke_{n}=(\nu_{n}-\nu_{n-1})(\Phi/\phi_{0})\;\;,
\ee
where $\Phi=AB$ is the flux through the island. Thus, for $K$
quasiholes, the flux through the island is given by
\be
\Phi=K (2n-1)\phi_{0}\;\;.
\label{flux}
\ee
Addition or removal of a single quasihole requires a flux change of
$(2n-1)\phi_{0}$ through the island, which gives the period
\be
\Delta \Phi =(2n-1)\phi_{0}\;\;.
\label{period}
\ee
When the island liquid is $\nu_{n+1}$ (and the channel liquid is $\nu_{n}$),
\be
Ke_{n}=(\nu_{n+1}-\nu_{n}) (\Phi/\phi_{0})\;\;,
\ee
and the period is given by
\be
\Delta \Phi=(2n+3)\phi_{0}\;\;.
\ee
In both cases, the periods are
in agreement with the general formula, Eq.~(\ref{tpd}).

It is instructive to consider this problem from yet another
perspective. We write pseudo wave functions in terms of the
coordinates of the quasiparticles, treating them as point particles
\cite {halperin}.
First consider the situation when the channel liquid is $\nu_{n}$ and
the island liquid is $\nu_{n-1}$.
Since the low-energy states contain quasiholes in the topmost level only
(i.e., related to holes only in the $n$th LL of $\Psi_{n}$), they fill
a lowest LL of their own. The most energetically favorable
situation is when they completely fill the LL. The wave function is
then
\be
\prod_{j<k}(\eta_{j}-\eta_{k})^{\theta}
\exp[-\frac{1}{4}\sum_{j=1}^K \frac
{|\eta_{j}|^2}{l_{n}^2}]\;\;,
\label{qhwf}
\ee
where $\eta_{j}$ denote the positions of the quasiholes, and
$l_{n}^2=\hbar c/e_{n} B$. The area of the island is given by
(neglecting irrelevant corrections of order unity) [14a]
\be
A=K \frac{\phi_{0}}{B} \frac{\theta}{(e_{n}/e)}\;\;.
\label{ratio}
\ee
With $\theta=\theta_{n}$, given by Eq.~(\ref{statistics}),
this is identical to Eq.~(\ref{flux}), and gives a period of
$(2n-1)\phi_{0}$.
In the other case, when the island liquid is $\nu_{n+1}$, we
write the quasielectron wave function \cite {halperin,khd1}
\be
\prod_{j<k}(\overline{\eta}_{j}-\overline{\eta}_{k})^{-\theta}
\exp[-\frac{1}{4}\sum_{j=1}^K \frac
{|\eta_{j}|^2}{l_{n}^2}]\;\;,
\label{qewf}
\ee
where now $\eta_{j}$ are the quasielectron coordinates.
In this case, one is tempted to choose the quasiparticle statistics
$\theta=\theta_{n}$. However, in
order for the quasielectron wave function to be regular as two
quasielectrons approach one another,
which is required by the hermiticity of the Hamiltonian \cite {khd},
we must choose the statistics to be \cite {ma}
\be
\theta =\theta_{n}-2= -\frac{2n+3}{2n+1}\;\;.
\ee
(The resulting quasielectron wave function can also be interpreted
as a FQHE liquid of quasielectrons of statistics $\theta_{n}$
\cite {halperin}.) The period from Eq.~(\ref{ratio}) is
$(2n+3)\phi_{0}$, as expected. From this perspective, the period
can be interpreted as a measure of the {\em ratio}
of the statistics to charge of the quasiparticles of the channel FQHE
liquid (see Eq.~\ref{ratio}).

We close with the following remarks.

(i) The above arguments actually show that for a consistent description
in terms of quasiparticles, they {\em must} be assigned fractional
statistics. Similar arguments had originally led Halperin to discover
that quasiparticles obey fractional statistics \cite {halperin}.

(ii) It is interesting to see how the BY result is obtained from the
perspective of the quasiparticles. This pertains to the situation when
charge is completely depleted from the island region. In the
CF theory, this relates to the IQHE state in which all $n$ LLs
are empty in the island region. In the quasihole language, $n$ LLs
of quasiholes are occupied. In analogy with the CF theory, the wave
function of this quasihole state is given by \cite
{ma}
\be
\prod_{j<k}(\eta_{j}-\eta_{k})^{\theta_{n}-1} \Psi_{n}\;\;.
\ee
The size of the droplet described by this wave function is such that
the flux through it is given by
\be
\Phi=\frac{K}{n}\frac{\phi_{0}}{B}\;\;.
\ee
In this case, the number of quasiholes increases in units of $n$
(since, whenever it is possible to add a quasihole in
one level, it is possible in other levels as well), and we recover the BY
period of $\phi_{0}$.

(iii) We have so far assumed that the $\nu=p/q$ FQHE liquid
in the island is
ideal. It is easy to see that the presence of a {\em fixed} number
of quasielectrons or quasiholes in this liquid will not alter the
period. Whenever a {\em new} quasiparticle is created, the periodic sequence
will suffer a phase
shift. The same will be true when there are lakes of other
FQHE liquids inside the island; the period will
remain $j\phi_{0}$ except when a new quasiparticle
is created in one of the lakes. Thus, in general, we
expect {\em finite} sequences of peaks in the longitudinal resistance
with the predicted spacing. The larger the amount of the $\nu=p/q$
fluid in the island, the longer will be the length of the
sequence.

(iv) We have neglected the Coulomb blockade effects \cite {lee},
which are expected to be small for sufficiently large islands. These
are also  well understood and may be subtracted out to reveal the
effects discussed here. We note that the periodicity of the effect does
not depend on the structure of the interface between the two FQHE
liquids, so long as it is narrow compared to the regions of the FQHE
liquids.

(v) Any $j\phi_{0}$ periodicity ($j\neq 1$) in the situation when
the island is completely depleted, as is presumably the case
in the experiment of Simmons {\em et al.} \cite {simmons},
{\em must} be a non-equilibrium effect \cite {kivelson}. This should be
experimentally testable.

In conclusion, we predict conditions under which an interference
between two FQHE liquids allows the observation of
an effective flux quantum, which is equal to
an odd integer multiple of the fundamental flux quantum.
The period depends on the quasiparticle charges of the two FQHE
liquids; in the case of two successive FQHE states of a sequence,
it can be also interpreted as
a measure of the ratio of the statistics to the charge of the
quasiparticles of the channel FQHE liquid.
This experiment should also serve as a probe
into the internal structure of the FQHE liquids.

We thank V.J. Goldman for discussions and comments.  This work
was supported in part by the National Science Foundation under
Grants Nos. DMR90-20637, DMR90-11803, and DMR92-20733.
We acknowledge the hospitality of the Aspen Center for Physics.

\pagebreak

\pagebreak

{\bf Figure Caption}:

\vspace{.5 cm}

Figure 1. (a) Standard Aharonov Bohm geometry. (b) Schematic drawing of the
proposed resonant tunneling experiment. The shaded area is the island
of $\nu=p/q$ FQHE liquid surrounded by the $\nu'=p'/q'$ FQHE liquid. The dashed
lines show the most probable tunneling paths.


\begin{thebibliography}{99}
\bibitem{ff} The filling factor at magnetic field $B$
is defined by $\nu=\rho\phi_{0}/B$ where $\rho$ is the electron
density, and $\phi_{0}=hc/e$.
\bibitem{general} In general, the charge of the quasihole of the
$\nu=p/q$ liquid is given by $e/q$ and the statistics is given by
$\theta=p'/q$ where $p'p=1$ mod $q$. This simple expression for the
statistics (which is, of course, in agreement with those of Refs.\cite
{halperin,su}) was given by G. Moore and N. Read, Nucl. Phys. B
{\bf 360}, 362 (1991).
\bibitem{semiclassical} We will assume that the number of
quasiparticles $K>>1$ so that order unity corrections can be neglected.
\end{thebibliography}
\end{document}